\documentclass[prd,showpacs,showkeys,floatfix,twocolumn,amsmath,amssymb,floatfix]{revtex4}
\usepackage{graphicx,color,dcolumn,booktabs,bm}
\usepackage{longtable,lscape}
\usepackage{amssymb}
\usepackage{indentfirst}
\usepackage{epsfig}
\newcommand{\beq}{\begin{equation}}
\newcommand{\eeq}{\end{equation}}
\newcommand{\bqa}{\begin{eqnarray}}
\newcommand{\eqa}{\end{eqnarray}}

\begin{document}

\title{Pentaquark states in a diquark-triquark model}

\author{Ruilin Zhu$^{1,3}$}
\email{rlzhu@sjtu.edu.cn}
\author{Cong-Feng Qiao$^{2,3}$\footnote{Corresponding author}}
\email{qiaocf@ucas.ac.cn}

\affiliation{
$^1$INPAC, Shanghai Key Laboratory for Particle Physics and Cosmology, Department of Physics and Astronomy,
 Shanghai Jiao Tong University, Shanghai 200240, P.R. China\\
$^2$School of Physics, University of Chinese Academy of
Sciences, Beijing 100049, P.R. China \\$^3$CAS Center for Excellence in Particle Physics, Institute of High Energy Physics, Chinese Academy of Sciences, Beijing 100049, P.R. China}

\begin{abstract}
The diquark-triquark model is used to explain charmonium-pentaquark states, i.e., $P_c(4380)$ and $P_c(4450)$, which were observed recently by the LHCb collaboration. For the first time, we investigate the properties of the color attractive configuration of a triquark and we define a nonlocal light cone distribution amplitude for pentaquark states, where both diquark and triquark are not pointlike, but they have nonzero size. We establish an effective diquark-triquark Hamiltonian based on spin-orbital interaction. According to the Hamiltonian, we show that the minimum mass splitting between $\frac{5}{2}^+$ and $\frac{3}{2}^-$ is around $100$~MeV, which may naturally solve the challenging problem of small mass splitting between $P_c(4450)$ and $P_c(4380)$. This helps to understand the peculiarities of $P_c(4380)$ with a broad decay width whereas $P_c(4450)$ has a narrow decay width. Based on the diquark-triquark model, we predict more pentaquark states, which will hopefully be measured in future experiments.

\end{abstract}

\maketitle

The hadron spectrum has played an important role in understanding the inner hadron structure and for testing various models of hadrons with fundamental freedom. The study of hadron physics is also crucial for understanding the dynamics of quark and strong interaction, according to quantum chromodynamics (QCD). Conventional hadrons can be understood well by using the naive constituent quark model, where a meson comprises two constituent quarks, $q\bar{q}^\prime$, while a baryon is constructed from three constituent quarks, $qq^\prime q^{\prime\prime}$, with all in a color singlet. This simple description has been highly successful in the past half century. However, the quark model and QCD do not include a rule that forbids the existence of other multiquark states~\cite{GellMann:1964nj}, such as tetraquark or pentaquark states. In contrast to the conventional meson and baryon, finding the multiquark state, also known as the exotic state, has been a goal of particle physicists for many years.

Recent developments in exotic heavy hadron research started with the discovery of $X(3872)$ by the Belle Collaboration in 2003~\cite{Choi:2003ue}, which is distinguished by its narrow decay width ($\Gamma<1.2${MeV}). Subsequently, a series of exotic states, $XYZ$, were determined experimentally, which are difficult to embed in the conventional meson and baryon spectra, and thus they have attracted much attention from both theoretical and experimental researchers (e.g., see ~\cite{Agashe:2014kda} and the references therein). Recently, the LHCb Collaboration observed two exotic structures in the $J/\psi p$ channel of $\Lambda_b$ decay, which they referred to as pentaquark-charmonium states, $P_c(4380)$ and $P_c(4450)$~\cite{Aaij:2015tga}. One of these two structures has a mass of $4380\pm8\pm29${MeV}, a width of $205\pm18\pm86${MeV}, and a preferred spin-parity assignment of $J^P=\frac{3}{2}^-$, whereas the other is narrow with a mass of $4449.8\pm1.7\pm2.5${MeV}, a width of $39\pm5\pm19${MeV}, and a preferred spin-parity assignment of $J^P=\frac{5}{2}^+$.

The binding mechanism associated with these newly observed structures is still unclear. Various interpretations can be assigned according to the following three types of models. (i) The meson-baryon molecular model~\cite{Karliner:2015ina,Chen:2015loa,He:2015cea,Chen:2015moa,Huang:2015uda,Roca:2015dva}, where $P_c(4380)$ and $P_c(4450)$ are treated as the $\Sigma_c \bar{D}^*$ and $\Sigma^*_c \bar{D}^*$ bound states, respectively, or their mixture. For this model, the energy spectrum has been evaluated using a chiral effective Lagrangian approach~\cite{Chen:2015loa,He:2015cea}, the QCD sum rules~\cite{Chen:2015moa}, the color-screen model~\cite{Huang:2015uda}, and the scattering amplitudes approach~\cite{Roca:2015dva}.
(ii) Diquark(triquark) interaction models, for which the diquark-diquark-antiquark model~\cite{Maiani:2015vwa,Anisovich:2015cia,Li:2015gta,Ghosh:2015ksa,Wang:2015epa}
and compact diquark-triquark model \cite{Lebed:2015tna} have been proposed. (iii) The kinematic effect. In this model, the appearance of the structures $P_c(4380)$ and $P_c(4450)$ is attributed to the kinematic effect \cite{Guo:2015umn,Liu:2015fea,Mikhasenko:2015vca} rather than the bound states.

In previous studies, theoretical predictions of pentaquark states in the charmonium energy region were made before the LHCb observations. Previous predictions of hidden charm pentaquarks were reported by~\cite{Wu:2010jy,Yang:2011wz}. The production and decay properties of the structures $P_c(4380)$ and $P_c(4450)$ were also investigated by
\cite{Maiani:2015iaa,Burns:2015dwa,Scoccola:2015nia,
Wang:2015jsa,Karliner:2015voa,Cheng:2015cca,Lu:2015fva}.

In this letter, we attribute the $P_c(4380)$ and $P_c(4450)$ states to possible diquark-triquark states with a $[cu][ud\bar{c}]$ configuration,
where both the diquark and triquark are loosely bound units, which is a generalization of the compact diquark $\delta$ and triquark $\bar{\theta}$ introduced by Brodsky, Hwang, and Lebed ~\cite{Lebed:2015tna,Brodsky:2014xia,Lebed:2015sxa}. The diquark interaction model was first employed by Jaffe and Wilczek ~\cite{Jaffe:2003sg}, while Karliner and Lipkin ~\cite{Karliner:2003dt} gave an interpretation of the unconfirmed pentaquark state $\Theta^+$. According to QCD, their analyses can be simply transferred to the heavy quark sector, where two quarks attract each other to form a diquark, and two quarks with an antiquark are also bound up to a triquark. In the following, we show that the small mass splitting between $P_c(4450)$ and $P_c(4380)$, and their peculiar decay widths can be understood using the diquark-triquark model.

According to group theory, the color group $SU(3)$ of a diquark can be represented either by a antitriplet or sextet in the decomposition of $\mathbf{3}\otimes \mathbf{3}=\mathbf{\bar{3}}\oplus \mathbf{6}$,
whereas a triquark may belong to one of the four different representations of $\mathbf{3}\otimes \mathbf{3} \otimes \mathbf{\bar{3}} =(\mathbf{\bar{3}}\oplus \mathbf{6})\otimes \mathbf{\bar{3}}=\mathbf{3}\oplus \mathbf{\bar{6}}\oplus \mathbf{3}\oplus \mathbf{15}$. It should be noted that in the one-gluon-exchange model,
the binding of the $q_1\bar{q}_2$ or $q_1q_2$ system depends solely on the quadratic Casimir $C_2(R)$ of the product color representation
R to which the quarks couple according to the discriminator
$I=\frac{1}{2}(C_2(R)-C_2(R_1)-C_2(R_2))$, where $R_i$ denotes the color representations of two quarks \cite{Brodsky:2014xia}; thus, we can immediately obtain the discriminators $I=\frac{1}{6}(-8,-4,+2,+1)$ for $R = (\mathbf{1},\mathbf{\bar{3}},\mathbf{6},\mathbf{8})$, respectively. When
$I$ is negative, the interaction force will be attractive, which is
somewhat analogous to the Coulomb force in QED. Thus, the only color attractive configuration of $q_1\bar{q}_2$ is in the color-singlet $\mathbf{1}$, whereas the color attractive configuration of $q_1q_2$ is in the color antitriplet $\mathbf{\bar{3}}$. In the one-gluon-exchange interaction, the attractive force strength in the color-singlet $q_1\bar{q}_2$ is two times that in the diquark $q_1q_2$. Without any loss of generality, the color structure of the triquark $q_3q_4\bar{q}_5$ can be taken as the product of a diquark $q_3q_4$ and an antiquark $\bar{q}_5$, and thus it can be decomposed as $(\mathbf{\bar{3}}\oplus \mathbf{6} )\otimes \mathbf{\bar{3}}=(\mathbf{\bar{3}}\otimes \mathbf{\bar{3}})\oplus (\mathbf{6}\otimes \mathbf{\bar{3}})=(\mathbf{3}\oplus \mathbf{\bar{6}})\oplus (\mathbf{3}\oplus \mathbf{15})$. Correspondingly, the discriminator $I=\frac{1}{6}(-4,+2,-5,+2)$ for $R=(\mathbf{3},\mathbf{\bar{6}},\mathbf{3},\mathbf{15})$, respectively. Obviously, there are two types of attractive color configurations for the triquark $q_3q_4\bar{q}_5$. One is in the color triplet $\mathbf{3}$ with $\bar{q}_5$ attracting $q_3 q_4$, where $q_3$ is repulsive to $q_4$, which is analogous to helium composed of a nucleus and two electrons. The other is also in the color triplet $\mathbf{3}$ with $\bar{q}_5$ attracting $q_3 q_4$, but $q_3$ is attractive to $q_4$, which is a peculiar interaction structure obtained from QCD. According to this analysis, we find that the diquark $q_1q_2$ in color configuration $\mathbf{\bar{3}}$ and the triquark $q_3q_4\bar{q}_5$ in color configuration $\mathbf{3}$ may form a color-singlet pentaquark state $q_1q_2 q_3 q_4 \bar{q}_5$.

Before starting the spectrum analysis, we first need to define a light cone distribution amplitude for the pentaquark $P_Q$ in terms of nonlocal quark fields
\begin{widetext}
\vskip -0.5cm
\begin{eqnarray}
 \phi (w_i){u}_\gamma &=& \int \frac{dz_1^-dz_2^-dz_3^-dz_4^-}{(2\pi)^4} e^{-ik^+(w_1z_1^-+w_2z_2^-+w_3z_3^-+w_4z_4^-) } \epsilon^{abc}\epsilon^{def}\epsilon^{cfg}\nonumber\\&&\times \langle P_Q(k)|Q^\mathrm{T}_i(z_1^-) L^\mathrm{T}_{ia}(z_1^-,0)\Gamma_\alpha q_j(z_2^-)L_{jb}(z_2^-,0) {q^\prime}^\mathrm{T}_l(z_3^-)L^\mathrm{T}_{ld}(z_3^-,0) \Gamma_\beta q^{\prime\prime}_s(z_4^-)L_{se}(z_4^-,0)\bar{Q^\prime}^g_\gamma(0)|0\rangle\ ,
 \label{eq:definition-gauge-invariant-LCDA}
\end{eqnarray}
\end{widetext}
\vskip -1.5cm
where $k$ is the momentum of the pentaquark and in the lightcone definition $k^+=(k^0+k^3)/\sqrt{2}$ and $k^-=(k^0-k^3)/\sqrt{2}$, $w_i$ is the quark momentum fraction and the spinor ${u}_\gamma$ denotes the heavy antiquark $\bar{Q^\prime}$ with momentum fraction of $w_{\bar{Q^\prime}} = 1-\sum_{i=1,4} w_i$, which is at rest at the space-time origin. The letters a--g, i, j, l, and s represent color indices. For prompt pentaquark production, the leading-twist contribution comes from the collinear conformal subset~\cite{Zhu:2015qoa}, where the gauge link can be expressed as
\begin{equation}
L(x,y) = P\,e^{ig\int_0^1 ds (x-y)_\mu G^\mu((x-y)s+y)}\ .
\end{equation}
In this case, the gluon field $G^\mu(x)\equiv G_\lambda^\mu (x) T^\lambda$ lies in the adjoint representation. It should be noted that the gauge links connect to the quark fields in the fundamental representation, which ensures that all of the colored quarks are transported to the space-time origin, and thus the pentaquark is well defined. For differences in the spin-parity of the diquark, we have $\Gamma_{\alpha,\beta} =  C,\ C\gamma_\mu,\ C\sigma_{\mu\nu}, \ C\gamma_5\gamma_\mu,\  C\gamma_5$, which correspond to the scalar, vector, tensor, pseudovector, and pseudoscalar, respectively. The charge conjugation matrix $C$ is defined as $C=i\gamma_2\gamma_0$ in the Pauli-Dirac representation. In the following, we focus only on the scalar and vector diquarks, which are referred to as ``good" and ``bad" diquarks by Jaffe, respectively~\cite{Jaffe:2004ph}.

The general QCD confining potential for the multiquarks reads \cite{DeRujula:1975ge}
\begin{equation}
V(\vec{r}_i) = L(\vec{r}_1,\vec{r}_2,\ldots)+\sum_{i>j}I\, \alpha_s S_{ij}\ ,
\end{equation}
where $L(\vec{r}_i)$ represents the universal binding interaction of quarks, $S_{ij}$ denotes two-body Coulomb and chromomagnetic interactions, and $I=-\frac{4}{3}$ and $-\frac{2}{3}$ denote the coefficients of single-gluon interactions in the quark-antiquark and quark-quark cases, respectively.

The effective Hamiltonian includes spin-spin interactions inside the diquark and triquark, as well as between them, the spin-orbital and purely orbital interactions, which may be expressed formally as
\begin{eqnarray}
 H&=&m_{\delta}+m_\theta+H^{\delta}_{SS} + H^{\bar{\theta}}_{SS}+H^{\delta\bar{\theta}}_{SS} + H_{SL}+H_{LL}\nonumber\\
 \label{eq:definition-hamiltonian}
\end{eqnarray}
\vskip -0.5cm
with
\vskip -0.5cm
\begin{eqnarray}
 H^\delta_{SS}&=&2(\kappa_{Qq})_{\bar{3}}(\mathbf{S}_Q\cdot \mathbf{S}_q),\nonumber\\
 H^{\bar{\theta}}_{SS}&=&2(\kappa_{q^\prime q^{\prime\prime}})_{\bar{3}}(\mathbf{S}_{q^\prime}\cdot \mathbf{S}_{q^{\prime\prime}})+2\kappa_{q^\prime \bar{Q^\prime}}(\mathbf{S}_{q^\prime}\cdot \mathbf{S}_{\bar{Q^\prime}})\nonumber\\
 &&+2\kappa_{ q^{\prime\prime}\bar{Q^\prime}} (\mathbf{S}_{q^{\prime\prime}}\cdot \mathbf{S}_{\bar{Q^\prime}}), \nonumber\\
 H^{\delta\bar{\theta}}_{SS} &=&2(\kappa_{Qq^\prime})_{\bar{3}}(\mathbf{S}_Q\cdot \mathbf{S}_{q^\prime}) +2(\kappa_{Qq^{\prime\prime}})_{\bar{3}}(\mathbf{S}_Q\cdot \mathbf{S}_{q^{\prime\prime}})\nonumber\\
 &&+2(\kappa_{q q^{\prime}})_{\bar{3}}(\mathbf{S}_{q}\cdot \mathbf{S}_{q^{\prime}})+ 2(\kappa_{qq^{\prime\prime}})_{\bar{3}}(\mathbf{S}_q\cdot \mathbf{S}_{q^{\prime\prime}}) \nonumber\\
 &&+2\kappa_{Q\bar{Q^\prime}} (\mathbf{S}_Q\cdot \mathbf{S}_{\bar{Q^\prime}})+ 2\kappa_{q\bar{Q^\prime}}(\mathbf{S}_q\cdot \mathbf{S}_{\bar{Q^\prime}}),
\nonumber\\
 H_{SL}&=&2 A_\delta (\mathbf{S}_\delta \cdot \mathbf{L}) +2 A_{\bar{\theta}}(\mathbf{S}_{\bar{\theta}}\cdot \mathbf{L}),\nonumber\\
 H_{LL}&=&B_Q \frac{L(L+1)}{2}\ ,
\label{eq:definition-hamiltonian2}
\end{eqnarray}
where $m_\delta$ and $m_{\theta}$ are the constituent masses of the diquark $[Qq]$ and triquark $[q^\prime q^{\prime\prime}\bar{Q^\prime}]$, respectively; $ H^\delta_{SS}$
and $H^{\bar{\theta}}_{SS}$ describe the spin-spin interactions inside the diquark and triquark, respectively; $H^{\delta\bar{\theta}}_{SS}$ describes the spin-spin interactions of quarks between the diquark and triquark; $H_{SL}$ and $H_{LL}$ correspond to the spin-orbital and purely orbital terms, respectively; $\mathbf{S}_{q^{(\prime,\prime\prime)}}$, $\mathbf{S}_{Q}$, and $\mathbf{S}_{\bar{Q}^{\prime}}$ are spin operators for the light quarks, heavy quark, and antiquark, respectively; $\mathbf{S}_{\delta}$ and $\mathbf{S}_{\bar{\theta}}$ are the spin operators for the diquark and triquark, respectively; $\mathbf{L}$ is the orbital angular momentum operator; $\kappa_{q_1\bar{q}_2}$ and $(\kappa_{q_1q^\prime_2})_{\bar{3}}$ are the spin-spin couplings for a quark-antiquark pair and diquark in the color antitriplet, respectively; and $A_{\delta(\bar{\theta})}$ and $B_Q$ are spin-orbit and orbit-orbit couplings, respectively.

For low energy pentaquark states with a quark content of $[Qq][q^\prime q^{\prime\prime}\bar{Q^\prime}]$, the orbital angular momenta are null, i.e., $L=0$. In particular, in the case of spin-parity $J^P=\frac{1}{2}^-$, there are five possible pentaquark states, i.e.,
\vskip -0.5cm
\begin{widetext}
\vskip -0.5cm
\begin{eqnarray}
|0_\delta;0_{\delta^\prime},\frac{1}{2}_{\bar{Q}^\prime}, \frac{1}{2}_{\bar{\theta}};\frac{1}{2}\rangle&=&\frac{1}{2}
\big[(\uparrow)_Q(\downarrow)_q-(\downarrow)_Q(\uparrow)_q \big]\big[(\uparrow)_{q^\prime}(\downarrow)_{q^{\prime\prime}}
-(\downarrow)_{q^{\prime}}(\uparrow)_{q^{\prime\prime}}\big] (\uparrow)_{\bar{Q}^\prime},\nonumber\\
|0_\delta;1_{\delta^\prime},\frac{1}{2}_{\bar{Q}^\prime}, \frac{1}{2}_{\bar{\theta}};\frac{1}{2}\rangle&=&\frac{1}{\sqrt{3}}
\big[(\uparrow)_Q(\downarrow)_q-(\downarrow)_Q(\uparrow)_q\big] \{(\uparrow)_{q^\prime}(\uparrow)_{q^{\prime\prime}}
(\downarrow)_{\bar{Q}^\prime}-\frac{1}{2} [(\uparrow)_{q^\prime}(\downarrow)_{q^{\prime\prime}}
+(\downarrow)_{q^{\prime}}(\uparrow)_{q^{\prime\prime}}] (\uparrow)_{\bar{Q}^\prime}\},\nonumber\\
|1_\delta;0_{\delta^\prime},\frac{1}{2}_{\bar{Q}^\prime}, \frac{1}{2}_{\bar{\theta}};\frac{1}{2}\rangle &=&\frac{1}{\sqrt{3}}
\big[(\uparrow)_{q^{\prime}}(\downarrow)_{q^{\prime\prime}}- (\downarrow)_{q^{\prime}}(\uparrow)_{q^{\prime\prime}}\big]
\{(\uparrow)_Q(\uparrow)_q
(\downarrow)_{\bar{Q}^\prime}-\frac{1}{2}[(\uparrow)_Q(\downarrow)_q
+(\downarrow)_Q(\uparrow)_q](\uparrow)_{\bar{Q}^\prime}\},\nonumber\\
 |1_\delta;1_{\delta^\prime},\frac{1}{2}_{\bar{Q}^\prime}, \frac{1}{2}_{\bar{\theta}};\frac{1}{2}\rangle&=&\frac{1}{3}
(\uparrow)_Q(\uparrow)_q\{[(\uparrow)_{q^\prime}(\downarrow)_{q^{\prime\prime}}
+(\downarrow)_{q^{\prime}}(\uparrow)_{q^{\prime\prime}}] (\downarrow)_{\bar{Q}^\prime}-
2(\downarrow)_{q^\prime}(\downarrow)_{q^{\prime\prime}}
(\uparrow)_{\bar{Q}^\prime}\}\nonumber\\
&&-\frac{1}{6}\big[(\uparrow)_Q(\downarrow)_q +(\downarrow)_Q(\uparrow)_q\big]\{
2(\uparrow)_{q^\prime}(\uparrow)_{q^{\prime\prime}}
(\downarrow)_{\bar{Q}^\prime}-[(\uparrow)_{q^\prime} (\downarrow)_{q^{\prime\prime}}
+(\downarrow)_{q^{\prime}}(\uparrow)_{q^{\prime\prime}}] (\uparrow)_{\bar{Q}^\prime}\},\nonumber\\
|1_\delta;1_{\delta^\prime},\frac{1}{2}_{\bar{Q}^\prime}, \frac{3}{2}_{\bar{\theta}};\frac{1}{2}\rangle&=&
\frac{1}{\sqrt{2}}(\downarrow)_Q(\downarrow)_q (\uparrow)_{q^{\prime}}(\uparrow)_{q^{\prime\prime}}
(\uparrow)_{\bar{Q}^\prime}+\frac{1}{3\sqrt{2}} (\uparrow)_Q(\uparrow)_q
\{[(\uparrow)_{q^\prime}(\downarrow)_{q^{\prime\prime}}
+(\downarrow)_{q^{\prime}}(\uparrow)_{q^{\prime\prime}}] (\downarrow)_{\bar{Q}^\prime}+
(\downarrow)_{q^\prime}(\downarrow)_{q^{\prime\prime}}
(\uparrow)_{\bar{Q}^\prime}\}\nonumber\\&&-\frac{1} {3\sqrt{2}}\big[(\uparrow)_Q(\downarrow)_q+(\downarrow)_Q
(\uparrow)_q\big]\{
(\uparrow)_{q^\prime}(\uparrow)_{q^{\prime\prime}}
(\downarrow)_{\bar{Q}^\prime}+[(\uparrow)_{q^\prime} (\downarrow)_{q^{\prime\prime}}
+(\downarrow)_{q^{\prime}} (\uparrow)_{q^{\prime\prime}}](\uparrow)_{\bar{Q}^\prime}\}\ ,
 \label{eq:definition-states1/2}
\end{eqnarray}
\vskip -0.2cm
\end{widetext}
\vskip -0.3cm
where we use the notation $|S_\delta; S_{\delta'}, S_{\bar{Q}'}, J_{\bar{\theta}}; J \rangle$
 for pentaquark states. Here $S_\delta$ and $J_{\bar{\theta}}$ denote the spins of the diquark $[Qq]$ and triquark $[q^\prime q^{\prime\prime}\bar{Q^\prime}]$, respectively; $S_{\delta^\prime}$ and $S_{\bar{Q}^\prime}$ denote the spins of the diquark and antiquark within the triquark $\bar{\theta}$, respectively; and $J$ is the total angular momentum of the pentaquark.
In the following, for simplicity, we focus only on the scalar and vector diquarks, i.e., $S_{\delta^{(\prime)}}=0,1$.

For $J^P=\frac{3}{2}^-$, there are four possible pentaquark states, i.e.,
\begin{widetext}
\vskip -0.5cm
\begin{eqnarray}
|0_\delta;1_{\delta^\prime},\frac{1}{2}_{\bar{Q}^\prime}, \frac{3}{2}_{\bar{\theta}};\frac{3}{2}\rangle=|\frac{3}{2}^-\rangle_1&=&
\frac{1}{\sqrt{2}}\big[(\uparrow)_Q(\downarrow)_q -(\downarrow)_Q(\uparrow)_q\big](\uparrow)_{q^\prime}
(\uparrow)_{q^{\prime\prime}}(\uparrow)_{\bar{Q}^\prime},\nonumber\\
|1_\delta;0_{\delta^\prime},\frac{1}{2}_{\bar{Q}^\prime}, \frac{1}{2}_{\bar{\theta}};\frac{3}{2}\rangle=|\frac{3}{2}^-\rangle_2&=&
\frac{1}{\sqrt{2}}\big[(\uparrow)_{q^\prime}(\downarrow)_{q^{\prime\prime}}
-(\downarrow)_{q^\prime}(\uparrow)_{q^{\prime\prime}}\big](\uparrow)_Q
(\uparrow)_q(\uparrow)_{\bar{Q}^\prime},\nonumber\\
|1_\delta;1_{\delta^\prime},\frac{1}{2}_{\bar{Q}^\prime}, \frac{1}{2}_{\bar{\theta}};\frac{3}{2}\rangle=|\frac{3}{2}^-\rangle_3&=&
\frac{1}{\sqrt{6}}(\uparrow)_Q
(\uparrow)_q\{2(\uparrow)_{q^\prime} (\uparrow)_{q^{\prime\prime}}(\downarrow)_{\bar{Q}^\prime}
-[(\uparrow)_{q^\prime}(\downarrow)_{q^{\prime\prime}}
+(\downarrow)_{q^\prime}(\uparrow)_{q^{\prime\prime}}] (\uparrow)_{\bar{Q}^\prime}\},\nonumber\\
|1_\delta;1_{\delta^\prime},\frac{1}{2}_{\bar{Q}^\prime}, \frac{3}{2}_{\bar{\theta}};\frac{3}{2}\rangle=|\frac{3}{2}^- \rangle_4&=&\sqrt{\frac{3}{10}}\big[(\uparrow)_Q(\downarrow)_q +(\downarrow)_Q(\uparrow)_q\big]
(\uparrow)_{q^\prime}
(\uparrow)_{q^{\prime\prime}}(\uparrow)_{\bar{Q}^\prime}
-\sqrt{\frac{2}{15}}(\uparrow)_Q
(\uparrow)_q\{(\uparrow)_{q^\prime} (\uparrow)_{q^{\prime\prime}}(\downarrow)_{\bar{Q}^\prime}
\nonumber\\&&+[(\uparrow)_{q^\prime}(\downarrow)_{q^{\prime\prime}}
+(\downarrow)_{q^\prime}(\uparrow)_{q^{\prime\prime}}] (\uparrow)_{\bar{Q}^\prime}\}\ .
\label{eq:definition-states3/2}
\end{eqnarray}
\end{widetext}

For $J^P=\frac{5}{2}^-$, only one pentaquark state exists, i.e.,
\begin{eqnarray}
|1_\delta;1_{\delta^\prime},\frac{1}{2}_{\bar{Q}^\prime}, \frac{3}{2}_{\bar{\theta}};\frac{5}{2}\rangle&=&(\uparrow)_Q
(\uparrow)_q(\uparrow)_{q^\prime}(\uparrow)_{q^{\prime\prime}} (\uparrow)_{\bar{Q}^\prime}\ .
\label{eq:definition-states5/2}
\end{eqnarray}

We now consider the specific situation where $Q^\prime=Q=c$, $q^\prime=q=u$ and $q^{\prime\prime}=d$, which means that the pentaquarks are comprised of $[cu][ud\bar{c}]$. Then, for the state where $J^P=\frac{5}{2}^-$, the mass eigenvalue reads
\begin{equation}
 M(\frac{5}{2}^-)=m_\delta+m_{\theta}+\frac{ \kappa _{c\bar{c}}}{2}+\frac{3 \left[\kappa _{q\bar{c}}+(\kappa _{cq})_{\bar{3}}+(\kappa _{qq})_{\bar{3}}\right]}{2}\ ,
\end{equation}
where the isospin symmetry is maintained with $u=d=q$ and the small isospin breaking effect is discussed later.

Under the basis vectors $|\frac{3}{2}^-\rangle_i$ defined in Eq. (\ref{eq:definition-states3/2}), the mass splitting matrix $\Delta M$ for $J^P=\frac{3}{2}^-$ may be obtained as
\begin{widetext}
$\left(
\begin{array}{cccc}
 \frac{1}{2}\left( 2 \kappa _{q\bar{c}}-3 (\kappa _{cq})_{\bar{3}}+(\kappa _{qq})_{\bar{3}}\right) & 0 & \frac{1}{ \sqrt{3}}\left( \kappa _{q\bar{c}}- \kappa _{c\bar{c}}+(\kappa _{cq})_{\bar{3}} -
   (\kappa _{qq})_{\bar{3}}\right) & \frac{\sqrt{15}}{6}  \left(\kappa _{c\bar{c}}-\kappa _{q\bar{c}}+2 (\kappa _{cq})_{\bar{3}}-2 (\kappa _{qq})_{\bar{3}}\right) \\
 0 & h& 0 & 0 \\
 \frac{1}{ \sqrt{3}}\left( \kappa _{q\bar{c}}- \kappa _{c\bar{c}}+(\kappa _{cq})_{\bar{3}}- (\kappa _{qq})_{\bar{3}}\right) & 0 & \frac{1}{6}\left(7 (\kappa
   _{cq})_{\bar{3}}+7 (\kappa _{qq})_{\bar{3}}-\kappa _{c\bar{c}}-13 \kappa _{q\bar{c}}\right) & \frac{\sqrt{5}}{3}  \left( \kappa _{c\bar{c}}+ \kappa _{q\bar{c}}- (\kappa _{cq})_{\bar{3}}- (\kappa
   _{qq})_{\bar{3}}\right) \\
 \frac{\sqrt{15}}{6}  \left(\kappa _{c\bar{c}}-\kappa _{q\bar{c}}+2 (\kappa _{cq})_{\bar{3}}-2 (\kappa _{qq})_{\bar{3}}\right) & 0 & \frac{\sqrt{5}}{3}  \left( \kappa
   _{c\bar{c}}+ \kappa _{q\bar{c}}- (\kappa _{cq})_{\bar{3}}- (\kappa _{qq})_{\bar{3}}\right) & \frac{1}{6} \left(4 \kappa
   _{q\bar{c}}-2 \kappa _{c\bar{c}}-(\kappa _{cq})_{\bar{3}}-(\kappa _{qq})_{\bar{3}}\right)
\end{array}
\right)$~
\vskip 0.2cm
\end{widetext}
with  $h=\frac{1}{2} \left(\kappa _{c\bar{c}}+\kappa _{cq}+\kappa _{q\bar{c}}-3 \kappa _{qq}\right)$. It should be noted that $|\frac{3}{2}^-\rangle_2$ does not mix with other states due to the isospin symmetry. In the following, we show that the $|\frac{3}{2}^-\rangle_2$ state is actually the lowest mass state in the $J^P=\frac{3}{2}^-$ family.

\begin{table}[thb]
\caption{\label{tab:spin-spin coupling} The spin-spin couplings for the color-singlet quark-antiquark and color-antitriplet quark-quark pairs, where $q$ denotes u and d quarks.}
\begin{center}
\begin{tabular}{ccccccccccc}
\hline\hline\\
 $\mathrm{Spin-spin~ couplings}$ &$q\bar{q}$&$s\bar{s}$&$s\bar{q}$&$c\bar{q}$&$c\bar{s}$&$c \bar{c}$&$b\bar{q}$&$b\bar{s}$&$b\bar{c}$&$b\bar{b}$
 \\
 $(\kappa _{ij})_{0}$(MeV)&315&121&195&70&72&59&23&23&20&36\\
\hline
 $\mathrm{Spin-spin~ couplings}$ &qq&ss&sq&cq&cs&bq&bs&bc&&
 \\
 $(\kappa _{ij})_{\bar{3}}$(MeV)&103&72&64&22&25&6.6&7.5&10&\\
 \\
\hline\hline
\end{tabular}
\end{center}
\end{table}

\begin{figure}[th]
\begin{center}\vskip -0.2cm
\includegraphics[width=0.45\textwidth]{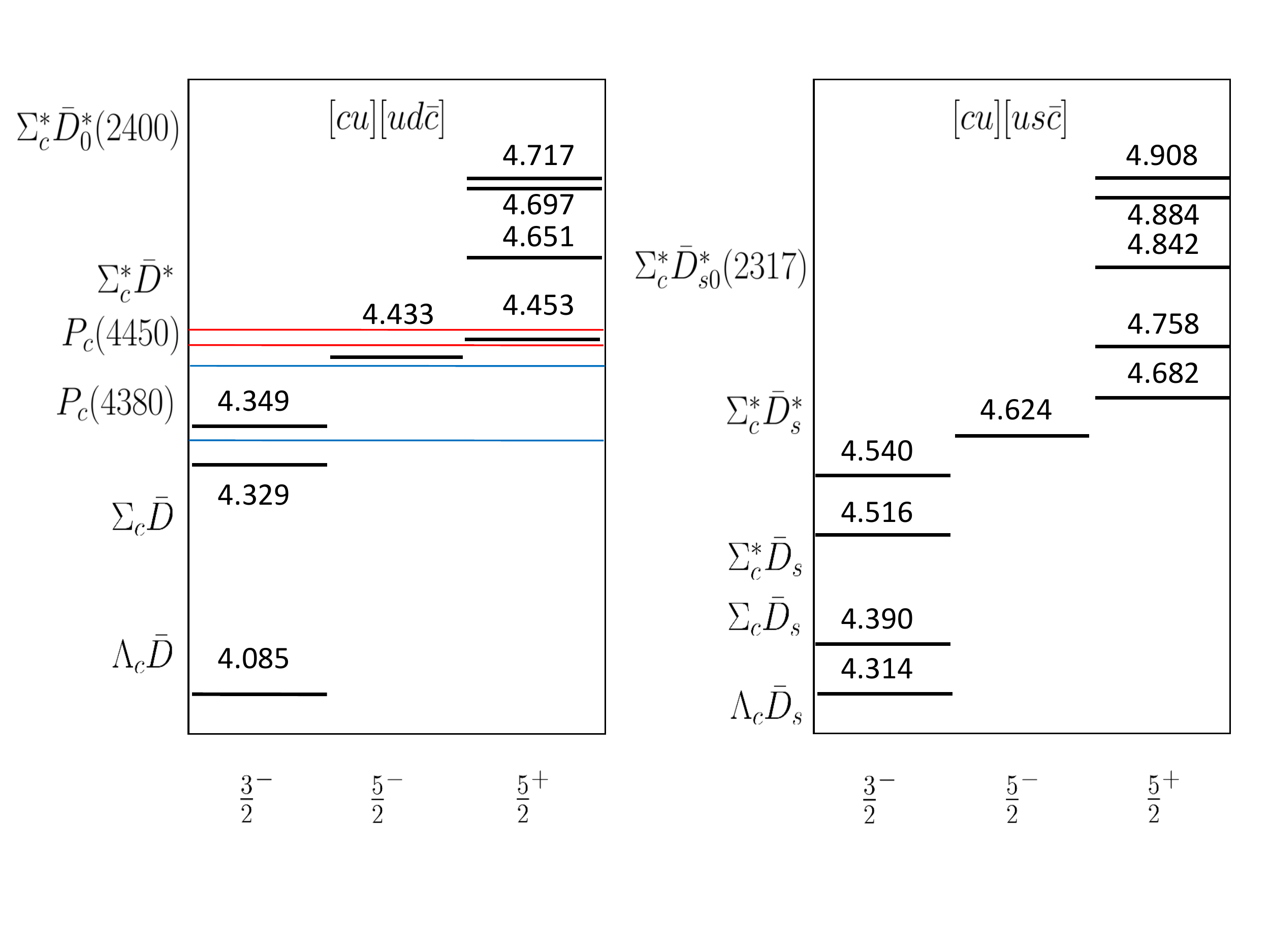}
\end{center}
\vskip -1.4cm
     \caption{The charmonium pentaquark spectra in GeV units with the quark constituents $[cu][ud\bar{c}]$ and $[cu][us\bar{c}]$. It should be noted that there are two degenerate states for $M=4.085${GeV} and $M=4.453${GeV} due to the isospin symmetry.}\label{Fig-spectrum1}
\end{figure}

\begin{figure}[th]
\begin{center}
\includegraphics[width=0.45\textwidth]{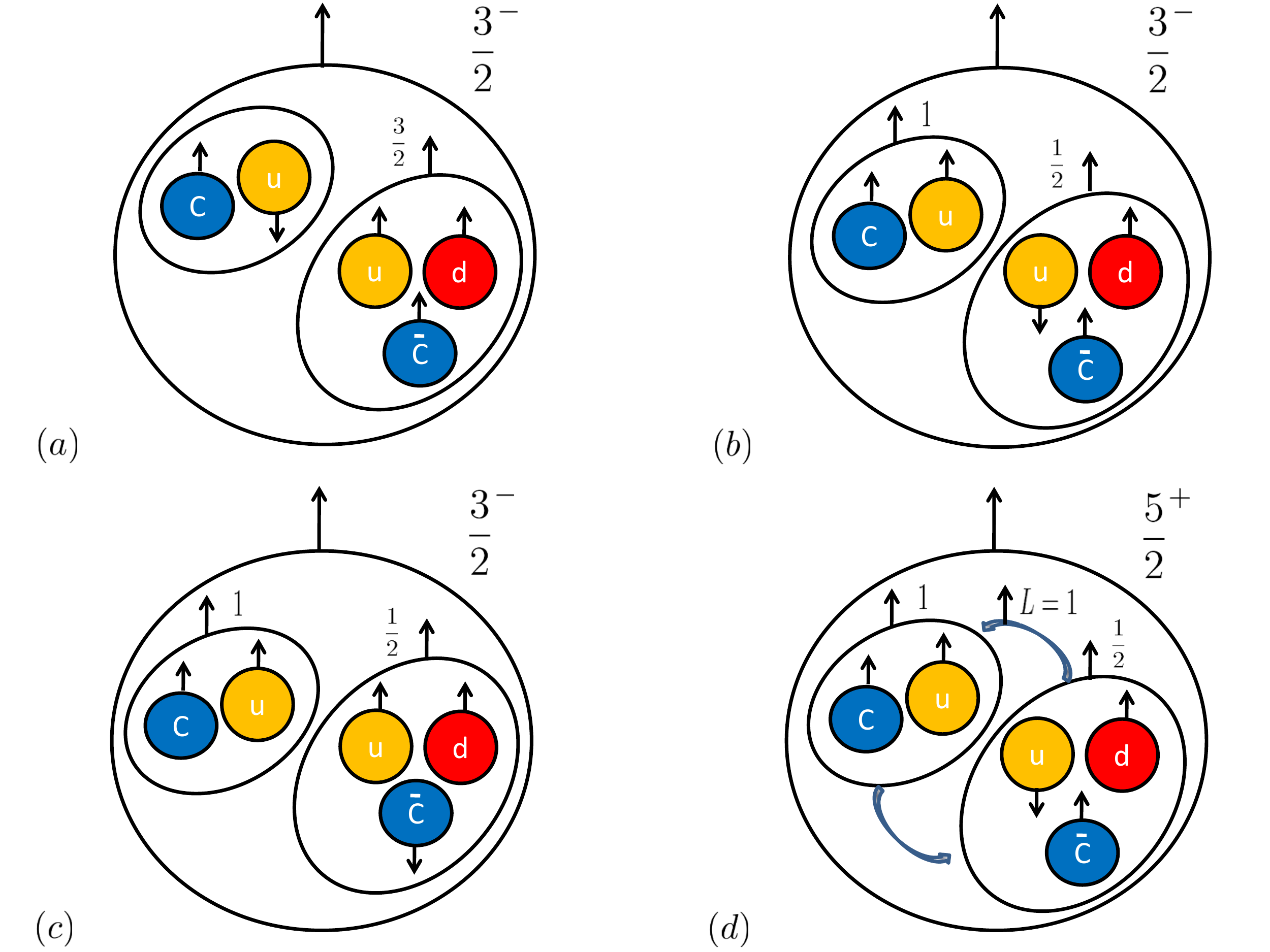}
\end{center}
    \vskip -0.5cm \caption{The possible diquark-triquark interpretation of the LHCb pentaquarks. $P_c(4380)$ is the heaviest state, a mixture of $|\frac{3}{2}^-\rangle_1$, $|\frac{3}{2}^-\rangle_3$, and $|\frac{3}{2}^-\rangle_4$, which can be simulated as the mixture of diagrams (a), (b), and (c). The $P_c(4450)$, corresponding to diagram (d) is an orbital excitation ($L=1$) of $|\frac{3}{2}^-\rangle_2$, which is the lowest mass state in the $J^P=\frac{5}{2}^+$ family. It should be noted that the inverse polarization cases for $(cu)_\delta$ in (a), $(ud)_{\bar{\theta}}$ in (b), and $(ud)_{\bar{\theta}}$ in (d) are implied.}\label{Fig-structure}
\end{figure}

For the first orbitally excited states, i.e., $L_{\delta\bar{\theta}}=1$, there are five pentaquark states with quantum number $J^P=\frac{5}{2}^+$.
According to Eq.(\ref{eq:definition-states3/2}), four of them are $|0_\delta;1_{\delta^\prime},\frac{1}{2}_{\bar{Q}^\prime}, \frac{3}{2}_{\bar{\theta}};\frac{3}{2}_{S},1_L,\frac{5}{2}_J\rangle$, $|1_\delta;0_{\delta^\prime},\frac{1}{2}_{\bar{Q}^\prime}, \frac{1}{2}_{\bar{\theta}};\frac{3}{2}_{S},1_L,\frac{5}{2}_J\rangle$, $|1_\delta;1_{\delta^\prime},\frac{1}{2}_{\bar{Q}^\prime}, \frac{1}{2}_{\bar{\theta}};\frac{3}{2}_{S},1_L,\frac{5}{2}_J\rangle$ and $|1_\delta;1_{\delta^\prime},\frac{1}{2}_{\bar{Q}^\prime}, \frac{3}{2}_{\bar{\theta}};\frac{3}{2}_{S},1_L,\frac{5}{2}_J\rangle$, and their corresponding Hamiltonians for spin-orbit and orbit-orbit interactions are $3A_Q+B_Q$, where the identical spin-orbit coupling $A_\delta=A_{\bar{\theta}}=A_Q$ is taken in Eq.~(\ref{eq:definition-hamiltonian2}) for simplicity. According to Eq.~(\ref{eq:definition-states5/2}), the fifth state with quantum number $J^P=\frac{5}{2}^+$ is $|1_\delta;1_{\delta^\prime}, \frac{1}{2}_{\bar{Q}^\prime}, \frac{3}{2}_{\bar{\theta}}; \frac{5}{2}_{S},1_L,\frac{5}{2}_J\rangle$,
and its corresponding Hamiltonians for spin-orbit and orbit-orbit interaction are $-2A_Q+B_Q$.

After inputting the spin-spin, spin-orbit, and orbit-orbit couplings, and masses of the quarks, we can readily obtain the pentaquark spectrum. For convenience, we give the spin-spin couplings in Table~\ref{tab:spin-spin coupling} \cite{Maiani:2004vq,Ali:2009es,Ali:2011ug,Ali:2014dva}, which are extracted from mesons, baryons, and the $XYZ$ spectra in the constituent quark model and diquark model. The expression $\kappa _{ij}=\frac{1}{4}(\kappa _{ij})_{0}$ for quark-antiquark coupling comes from the one gluon exchange model.

\begin{figure}[th]
\begin{center}
\includegraphics[width=0.45\textwidth]{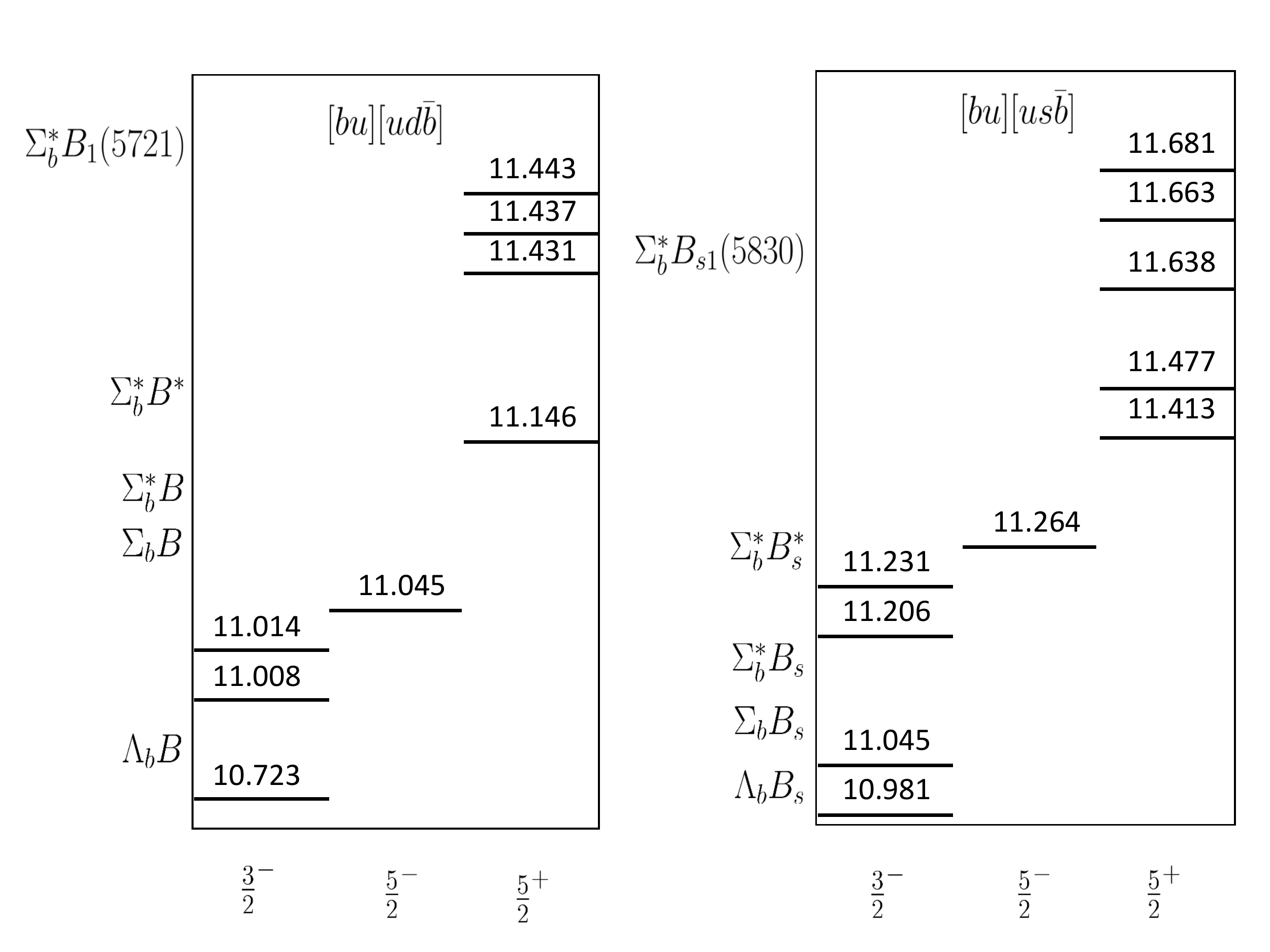}
\end{center}
\vskip -0.8cm
     \caption{The bottomonium pentaquark spectra with quark constituents $[bu][ud\bar{b}]$ and $[bu][us\bar{b}]$. It should be noted that there are two degenerate states, where $M = 10.723$ {GeV} and $M = 11.146$ {GeV}.}\label{Fig-spectrum2}
\end{figure}

The masses of diquarks $[cq]$ and $[bq]$ are extracted from $X(3872)$ with $J^{PC}=1^{++}$ and $Y_b(10890)$ with $J^{PC}=1^{--}$
in the diquark model, respectively. We find that $m_{[cq]}=1.932${GeV} and $m_{[bq]}=5.249${GeV}. In the numerical study of the pentaquark spectrum, the input quark masses are $m_q=305\mathrm{MeV}$, $m_s=490\mathrm{MeV}$, $m_c=1.670 \mathrm{GeV}$, and $m_b=5.008\mathrm{GeV}$ \cite{Ali:2009es,Ali:2011ug}. The spin-orbit coupling $A_Q$ takes $30\mathrm{MeV}\ \mathrm{and}\ 5${MeV} for $c$ and $b$ quarks, respectively; and the orbit-orbit coupling $B_Q$ takes $278\mathrm{MeV}$ and $408${MeV} for $c$ and $b$ quarks, respectively \cite{Ali:2014dva,Hambrock:2011qga,Maiani:2014aja}.
For triquark $\bar{\theta}[ud\bar{c}]$, the approximate relation $m_\theta\simeq m_c+2m_q=2.280${GeV} is employed.
The charmonium pentaquark spectra with quantum number $J^P=
\frac{3}{2}^-,\frac{5}{2}^-$, and $\frac{5}{2}^+$ are
depicted in Fig.~\ref{Fig-spectrum1}.

We find that among the many predicted pentaquark states, that with a mass of 4.349{GeV} probably corresponds to the LHCb $P_c(4380)$ state and that with a mass of 4.453{GeV} to the $P_c(4450)$. The minimum mass splitting between $\frac{5}{2}^+$ and $\frac{3}{2}^-$ states is about 100{MeV}, which explains the experimental measurements well. In general, it should be noted that the diquark-triquark model may give a large binding energy compared with the molecular model. Furthermore, we may also conclude that $P_c(4380)$ should not have the quantum number $J^P=\frac{5}{2}^+$ or $J^P=\frac{5}{2}^-$ by referring to Fig.~\ref{Fig-spectrum1}.
This is consistent with the LHCb measurement, where the
best fitting result shows that $P_c(4380)$ and $P_c(4450)$
probably have quantum numbers of $\frac{3}{2}^-$ and $\frac{5}{2}^+$, respectively, rather than
$J^P=(\frac{5}{2}^+, \frac{3}{2}^-)$ or $J^P=(\frac{3}{2}^+, \frac{5}{2}^-)$. The possible inner structures of $P_c(4380)$
and $P_c(4450)$ are depicted in Fig.~\ref{Fig-structure}, which indicate that $P_c(4450)$ is the lowest energy state in the $J^P=\frac{5}{2}^+$ family, whereas $P_c(4380)$ is the heaviest in the $J^P=\frac{3}{2}^-$ family. If we do not consider the
higher orbital excitations, $L\geq 2$, we can provide a qualitative explanation for why $P_c(4450)$ has a narrow width of 39{MeV} and $P_c(4380)$ has a broad width of 205{MeV}. The $P_c(4380)-P_c(4450)$ system is analogous to the $\Sigma(1940)-\Sigma(1915)$ system, where the $\Sigma(1940)$ is an excited state in the $J^P=\frac{3}{2}^-$ family and it has a broad decay width of about $220${MeV}, whereas $\Sigma(1915)$ is the lowest energy state with $J^P=\frac{5}{2}^+$ and it has a width of about $120${MeV}. In addition, in
our analysis, since $P_c(4380)$ is a mixture of $|\frac{3}{2}^-\rangle_1$, $|\frac{3}{2}^-\rangle_3$, and $|\frac{3}{2}^-\rangle_4$ according to Eq.~(\ref{eq:definition-states3/2}), then it naively has more decay channels and hence a broad decay width.

We show the bottomonium pentaquark spectra in Fig.~\ref{Fig-spectrum2}. It should be noted that many novel pentaquark states are predicted in Figs.~\ref{Fig-spectrum1} and \ref{Fig-spectrum2} according to the diquark-triquark model. In particular, those with relatively narrow decay widths and large masses are more likely to be detected in experiments. For the charmonium pentaquark with $S=0$, the predicted state with a mass of 4.329{GeV} and $J^P=\frac{3}{2}^-$ may be reconstructed through the $J/\psi p$ invariant mass distribution in the $\Lambda_b^0\to J/\psi p K^-$ decay channel, in a similar manner to the measurement of $P_c(4380)$ and $P_c(4450)$. Furthermore, the state with mass of 4.433{GeV} and $J^P=\frac{5}{2}^-$ may be reconstructed through the $J/\psi \Delta^+$ invariant mass distribution in the $\Lambda_b^0\to J/\psi \Delta^+ K^-$ decay channel. For the state with mass $4.085${GeV}, the decay channels with $J/\psi p$ in the final states would be difficult to measure due to the small phase space. For states over $4.6${GeV} in the left diagram in Fig.~\ref{Fig-spectrum1}, their masses exceed the $\Sigma_c^* \bar{D}^*$ threshold, which means that more decay channels are open, so they would be relatively difficult to measure in experiments.

The reconstruction of charmonium pentaquark states with strange number $S=-1$ is very similar to that of $P_c$ states with $S=0$. Given this feature, we suggest that the predicted charmonium pentaquark states ($S=-1$) with masses of $4.516${GeV}, $4.540${GeV}, and $4.682${GeV} may be detected through the $\Xi_b^0 \to J/\psi\Sigma^+ K^-$, $J/\psi \Sigma^0\bar{K}^0$ and $\Lambda_b^0 \to J/\psi\Lambda \phi$, $J/\psi\Sigma^0\phi$ channels, whereas the state with a mass of 4.624{GeV}, spin-parity $J^P=\frac{5}{2}^-$, and strange number $S=-1$ may be reconstructed through the $J/\psi \Sigma^+(1385)$ spectrum in the $\Xi_b^0\to J/\psi \Sigma^+(1385) K^-$ decay channel. The other states shown in the right diagram in Fig.~\ref{Fig-spectrum1} would be relatively difficult to measure due to either the small phase space or the possibly broad decay width.

Overall, the $P_c$ states tend to exhibit themselves in beauty-baryon decays, such as $\Lambda_b^0 \to J/\psi p K^-$, $J/\psi \Delta K^-$, $J/\psi n \bar{K}^0$, $J/\psi\Lambda \phi$, $J/\psi\Sigma^0\phi$, and $\Xi_b^0 \to J/\psi\Sigma^+ K^-$, $J/\psi \Sigma^0\bar{K}^0$. A neutral $P_c$ with $S=-1$ was predicted by\cite{Chen:2015sxa} as measurable through the $\Xi_b^- \to P_c K^- \to J/\psi \Sigma^+K^-$ process. It should be noted that the prompt production process also needs to be considered in the study of the pentaquark, e.g., through $p+p~(\gamma+p)\to (J/\psi p , J/\psi \Sigma^+, J/\psi \Lambda,\Upsilon p , \Upsilon \Sigma^+, \Upsilon \Lambda)+X$. For the $P_c$ states, those with a configuration of $[cu][ud\bar{c}]$ may have decay modes of $P_c\to J/\psi p$, $J/\psi \Delta$, $\Lambda_c\bar{D}$, and those with a configuration of $[cu][us\bar{c}]$ may decay through $P_c\to J/\psi \Sigma^+$, $\Lambda_c\bar{D}_s$. For $P_b$ states, those with a configuration of $[bu][ud\bar{b}]$ may decay through
$P_b\to \Upsilon p$, $\Upsilon \Delta$, $\Lambda_b B$, and those with a configuration of $[bu][us\bar{b}]$ might tend to decay through
$P_b\to \Upsilon \Sigma^+$, $\Lambda_b B_s$. Moreover, we may  reanalyze the strange pentaquark $\Theta^+$ through $\Lambda_c^+\to \Theta^+ \bar{K}^0\to(K^+ n, \bar{K}^0 p)\bar{K}^0$, in addition to $\Lambda_c^+ \to P_s \pi^0\to\phi p\pi^0$ and $\Xi_c^+ \to P_s\bar{K}^0\to\phi p\bar{K}^0$ processes \cite{Lebed:2015dca}.

The reconstruction of the bottomonium pentaquark is tedious because no hadron can decay directly to yield it. Thus, searching for the bottomonium pentaquark must rely on its prompt production in hadron-hadron collisions or lepton-hadron deep inelastic scattering processes.

In conclusion, we demonstrated that the diquark-triquark model may provide a good explanation of the pentaquarks discovered by the LHCb Collaboration. The small mass splitting between $P_c(4450)$ and $P_c(4380)$, and their special decay widths can be understood well using this model. We predicted more heavy pentaquark states, which may be confirmed by LHCb, JLab, or Belle-II experiments. Thus, the observation or non-observation of these states will facilitate the judgment of the diquark-triquark model. We also consider that it would be useful to analyze the $J/\psi \Sigma^+$ ($J/\psi \Lambda$) invariant mass spectrum in experiments such as LHCb, near $4.682${GeV} in $\Xi_b^0 \to J/\psi \Sigma^+K^- $ and $\Xi_b^- \to J/\psi\Lambda K^-$ decay channels, where charged and neutral charmonium-pentaquarks with $J^P=\frac{5}{2}^+$ and $S=-1$ may exist.

{\bf Acknowledgements:} We acknowledge useful discussions with Xiangdong Ji, Richard F. Lebed, Wei Wang, Timothy J. Burns, Xiao-Gang He, Lie-Wen Chen, Haijun Yang, Liming Zhang, and Kaijia Sun. This study was supported in part by a key laboratory grant from the Office of Science and Technology,
Shanghai Municipal Government (No. 11DZ2260700 and No. 15DZ2272100), by the Ministry of Science and Technology of the People's Republic of China (2015CB856703), and by the National Natural Science Foundation of
China under Grants No. 11175249 and No. 11375200.

\end{document}